%% file: Matousek_SPIN18_arXiv.tex
\documentclass[a4paper]{article}
\usepackage{graphicx}

\pdfoutput=1    

\def \ShortTitle#1{\def\@ShortTitle{#1}}
\def \FullConference#1{\def\@FullConference{#1}}

\renewcommand{\abstract}[1]
	{\gdef\abstract{#1}}

\input{header}

\newcommand{\speaker}[1]{#1}
\newcommand{\email}[1]{{\normalsize \texttt{#1}}}
\date{}

\begin{document}

\maketitle

\vspace{-15pt}
\parbox{.9\textwidth}{	
    \begin{center}
        Presented at \textit{\let\\=\relax \@FullConference}
    \end{center}
    }

\centerline{
    \parbox{.9\textwidth}{	
    \begin{center}
        {\bf Abstract}
    \end{center}
    \vspace{-5pt}
    \abstract}
    }

\input{text}

\input{Matousek_SPIN18.bbl}
\end{document}

%% file: header.tex
\usepackage{amsmath}
\usepackage{bm}		
\newcommand{\beq}{\begin{equation}}
\newcommand{\eeq}{\end{equation}}
\newcommand{\dif}[1]{\mathrm{d} #1 \,}

\newcommand{\GeVc}{GeV/$c$}
\newcommand{\GeVcc}{GeV/$c^2$}
\newcommand{\rarr}{\rightarrow}
\newcommand{\phiH}{\phi_\mathrm{h}}
\newcommand{\phiS}{\phi_\mathrm{S}}
\newcommand{\kT}{\vec{k_\mathrm{T}}}
\newcommand{\kTsq}{k_\mathrm{T}^2}

\newcommand{\PhTscal}{P_{h\mathrm{T}}}

\newcommand{\qT}{\vec{q_\mathrm{T}}}
\newcommand{\qTscal}{q_\mathrm{T}}
\newcommand{\qTscalt}{$q_\mathrm{T}$}
\newcommand{\kaT}{\vec{k_{\pi\mathrm{T}}}}

\newcommand{\kbT}{\vec{k_{\mathrm{pT}}}}

\newcommand{\fSiv}[1]{f_{1\mathrm{T}}^{\perp #1}}
\newcommand{\fSivb}[1]{f_{1\mathrm{T,p}}^{\perp #1}}
\newcommand{\F}[2]{F_\mathrm{#1}^{#2}}
\newcommand{\A}[2]{A_\mathrm{#1}^{#2}}
\newcommand{\qqswap}{(q\leftrightarrow\bar{q})}
\newcommand{\refeq}[1]{Eq.~(\ref{#1})}
\newcommand{\mr}[1]{\mathrm{#1}}
\renewcommand{\vec}[1]{\bm{#1}}

\title{Weighted transverse spin asymmetries in 2015 COMPASS Drell--Yan data}

\ShortTitle{Weighted TSAs in COMPASS Drell--Yan data}

\author{\speaker{J.~Matou\v{s}ek}, on behalf of the COMPASS Collaboration\\
        University and INFN Trieste, Italy\\
        E-mail: \email{jan.matousek@ts.infn.it}}

\abstract{Muon pairs created in the Drell--Yan process of a negative pion beam of 190\,\GeVc\ momentum impinging on a transversely polarised NH$_3$ target were detected with the COMPASS spectrometer at CERN in 2015. In addition to the extraction of the transverse spin asymmetries (TSAs) from these data, a complementary analysis of the TSAs weighted by powers of the dimuon momentum \qTscalt\ has been performed and is presented in this paper. In the transverse momentum dependent parton distribution functions (TMD PDFs) formalism, the \qTscalt-weighted TSAs can be interpreted in terms of products of the TMD PDFs of the beam pion and of the transversely polarised target proton, unlike the conventional TSAs, which are interpreted as their convolutions. This allowed us to make a straightforward comparison with the expectation based on the weighted Sivers asymmetry measured in the SIDIS process. In a similar way, the Boer--Mulders function of the pion and proton can be obtained as well.}

\FullConference{23rd International Spin Physics Symposium - SPIN2018 -\\
		10-14 September, 2018\\
		Ferrara, Italy}

%% file: text.tex
\section{Introduction}
The hadron structure is commonly described at leading twist by eight transverse momentum dependent (TMD) parton distribution functions (PDFs), which depend on the fraction $x$ of the hadron momentum carried by the parton and on the transverse component of the parton momentum $\kT$. They encode all possible correlations between the hadron spin, parton spin and $\kT$. They have been probed in semi-inclusive deep-inelastic scattering (SIDIS), in the Drell--Yan process (DY) and in hadron--hadron collisions. COMPASS played an important role in the first process and recently also in the second.

The idea of weighted transverse spin azimuthal asymmetries (TSAs) appeared first in the context of SIDIS\,\cite{kotzinian:1996,boer:1998}, but it is used in DY as well\,\cite{sissakian:2005b,wang:2017}. Their advantage is that the convolutions of the TMD PDFs, which are present in the standard interpretation of the SIDIS or DY cross-sections 
and consequently in the standard TSAs, are replaced by products of transverse moments of the TMD PDFs. COMPASS has recently complemented its results on the standard TSAs in SIDIS by the weighted Sivers asymmetries in SIDIS\,\cite{compass:2018weighted}, which offer a straightforward interpretation and insight into the validity of the Gaussian model of the $\kT$-depenence of the Sivers function. The TSAs in Drell--Yan from 2015 data\,\cite{compass:2017dy} were also accompanied by the corresponding weighted TSAs\,\cite{matousek:2017}, which are discussed here. New results are expected soon from the DY data collected in 2018.
\section{Transverse momentum weighted asymmetries in Drell--Yan process}
\label{sec:measurement}
We studied the DY reaction $\pi^- \mr{p}^\uparrow \rarr \mu^- \mu^+ X$ with a 190~\GeVc\ $\pi^-$ beam and a transversely polarised proton target. At leading order (LO), the reaction proceeds via annihilation of a quark--antiquark pair into a virtual photon. The cross-section consists of five terms, each posessing an orthogonal modulation in $\phi$ or $\phiS$ -- the azimuthal angles of the muon momentum and of the target polarisation vector, respectively\,\cite{arnold:2008}. Also, each term contains a structure function $\F{U/T}{X}$, which can be written as a flavour sum of convolutions of TMD PDFs over the intrinsic momenta of the two colliding quarks $\kaT$ and $\kbT$. The standard TSAs are ratios of the convolutions $\A{U/T}{X} = \F{U/T}{X} / \F{U}{1}$.

The convolutions can be solved assuming a certain $\kT$-dependence of the TMD PDFs; often a Gaussian form is used. The weighted asymmetries, on the other hand, are exploiting the fact that if one integrates the structure functions over $\qT = \kaT + \kbT$ with cleverly chosen weights $W_{X}$, the convolutions are solved trivially. The \qTscalt-weighted TSAs are then
$\A{T}{X W_X} = \int \dif{^2\qT} W_X \, \F{T}{X} / \int \dif{^2\qT} \F{U}{1}$. In particular, the three ones accessible in the $\pi\mr{p}^\uparrow$ DY interaction are 
\begin{align}
    \label{eq:siv}
    \A{T}{\sin\phiS \frac{\qTscal}{M_\mr{p}}}(x_\pi,x_N)
			&= - 2 \frac{\sum_q e_q^2 \bigl[ f_{1,\pi^-}^{\bar{q}}(x_\pi) \, 
						\fSivb{(1)q}(x_N) + \qqswap \bigr]}
				 {\sum_q e_q^2 
					\bigl[ f_{1,\pi^-}^{\bar{q}}(x_\pi)\, 
						   f_{1,\mr{p}}^q(x_N) + \qqswap \bigr]}
    \\
    \label{eq:transv}
	\A{T}{\sin(2\phi-\phiS) \frac{\qTscal}{M_\pi}}(x_\pi,x_N)
			&= - 2 \frac{ \sum_q e_q^2 \bigl[ 
							h_{1,\pi^-}^{\perp (1) \bar{q}}(x_\pi) \, 
							h_{1,\mr{p}}^q (x_N) 
						+ \qqswap \bigr]}
				 {\sum_q e_q^2 
					\bigl[ f_{1,\pi^-}^{\bar{q}}(x_\pi) f_{1,\mr{p}}^q(x_N) + \qqswap \bigr]}
    \\
    \label{eq:pretz}
	\A{T}{\sin(2\phi+\phiS) \frac{\qTscal^3}{2M_\pi M_\mr{p}^2}}(x_\pi,x_N)
			&= - 2 \frac{ \sum_q e_q^2 \bigl[ 
							h_{1,\pi^-}^{\perp (1) \bar{q}}(x_\pi) \, 
							h_{1\mr{T,p}}^{\perp (2) q} (x_N) 
						+ \qqswap \bigr]}
				 {\sum_q e_q^2 
					\bigl[ f_{1,\pi^-}^{\bar{q}}(x_\pi) f_{1,\mr{p}}^q(x_N) + \qqswap \bigr]}
\end{align}
where the sums run over quarks and antiquarks $q$; $e_q$ are fractional electric charges; $M_{\pi,\mr{p}}$ are the pion and proton masses; and $f^{(n)}$ or $h^{(n)}$ are the $n$-th $\kTsq$-moments of the TMD PDFs, 
\beq
	\label{eq:wpdf}
	f^{(n)}(x) = \int \dif{^2\kT}
					\biggl( \frac{\kTsq}{2M^2} \biggr)^n f(x,\kTsq).
\eeq

Each weighted TSA is obtained from the data by a fit of the so-called `modified double ratio' $R(\Phi) \propto \A{T}{\sin\Phi\,W}\,\sin\Phi$, where $\Phi = \phiS, 2\phi-\phiS, 2\phi+\phiS$, which is constructed from event counts and sums of event weights coming from two oppositely polarised target cells and from two `sub-periods' of data-taking. The polarisation of the cells is reversed between the sub-periods. The ratio is calculated in eight bins in $\Phi$, so the azimuthal acceptance $a(\Phi)$ is cancelled\,\cite{compass:2018weighted}. For the details of the target geometry, data-taking strategy and the anaysis we refer to the published paper\,\cite{compass:2017dy}.

\begin{figure}[t]
    \includegraphics[width=\textwidth]{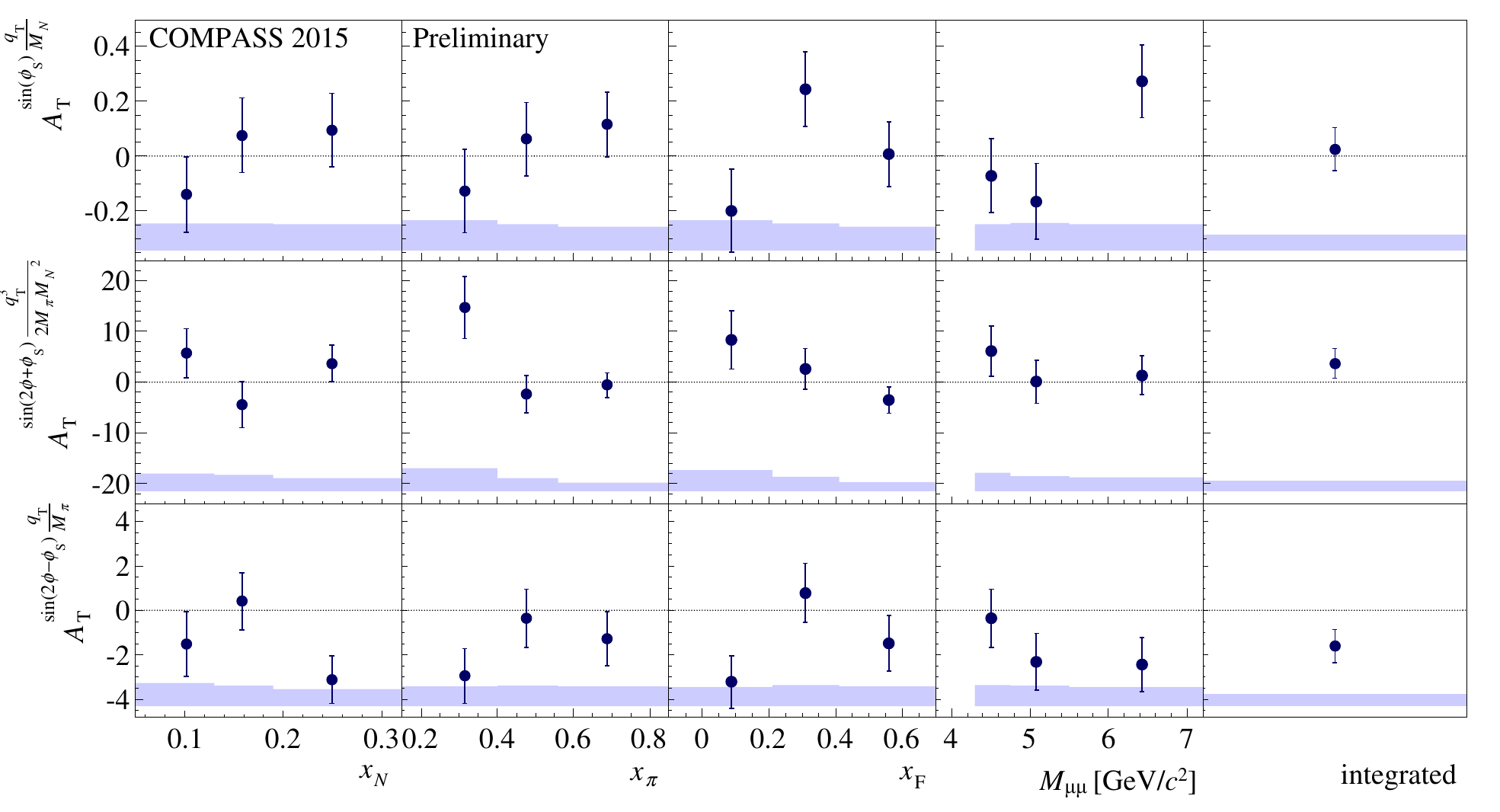}
    \caption{\label{fig:wasym} The \qTscalt-weighted TSAs. The systematic uncertainty is denoted by blue bands. Normalisation uncertainties from target polarisation (5\,\%) and dilution factor calculation (8\,\%) are not shown.}
\end{figure}

The data sample was collected in 2015 and the applied selections were almost the same as for the standard TSAs\,\cite{compass:2017dy}, in particular the same invariant mass range $M \in [4.3, 8.5]$\,\GeVcc\ is used. The only difference is the absence of cuts on \qTscalt. Instead, we applied a limit on the individual muon transverse momentum $l_\mr{T} < 7$\,\GeVc. The selected sample consists of about 39\,000 dimuons. The target composition (solid NH$_3$ with the H nuclei polarised, in He bath) is taken into account by a dilution factor, so the TSAs refer to proton. The results are shown in Fig.~\ref{fig:wasym}.

The background from other processes was estimated to be about~4\,\%\,\cite{compass:2017dy}. Several possible systematic effects were studied. The impact of a possible imperfection of the acceptance cancellation was estimated by the method of `false asymmetries'. They are calculated from events with the target cell of origin or the sub-period changed in such a way that the physics asymmetries cancel. The false asymmetries give the largest contribution to the systematic uncertainty, which turned out to be 0.7 times the statistical one. In addition, there are normalisation uncertainties from target polarisation (5\,\%) and dilution factor calculation (8\,\%).

\section{Transverse momentum weighted Sivers asymmetry in SIDIS and Drell--Yan}
\label{sec:siversproj}
The weighted asymmetries offer a unique and straightforward way to compare the results obtained by COMPASS in SIDIS and in DY. In SIDIS, the $\PhTscal/z$-weighted Sivers asymmetry\,\cite{compass:2018weighted} in the production of charged hadrons $h$ with available energy fraction $z>0.2$ can be written as
\beq
	A_{\mr{UT,T},h^\pm}^{\sin(\phiH-\phiS) \frac{\PhTscal}{zM}}(x)
			= 2\frac{ \frac{4}{9} \, \fSiv{(1)\mr{u}}(x,Q^2) \, 
						    \tilde{D}_{1,\mr{u}}^{h^\pm}(Q^2)
				     + \frac{1}{9} \, \fSiv{(1)\mr{d}}(x,Q^2) \, 
							\tilde{D}_{1,\mr{d}}^{h^\pm}(Q^2)}
			 {\sum_{q=\mr{u},\mr{d},\mr{s},
					\mr{\bar{u}},\mr{\bar{d}},\mr{\bar{s}}} 
					e_q^2 \, f_1^q(x,Q^2) \tilde{D}_{1,q}^{h^\pm}(Q^2)},
\eeq
where we use standard SIDIS variables and
$\tilde{D}_{1,q}^{h^\pm}(Q^2) = \int_{0.2}^1 \dif{z}\,D_{1,q}^{h^\pm}(z,Q^2)$
is the fragmentation function (FF) of a quark $q$ into the hadron $h$, integrated over $z$. We consider only u, d and s quarks in the proton, and we assume the Sivers function of sea quarks to be zero. 

\begin{figure}[t]
\begin{minipage}{0.48\textwidth}
\includegraphics[width=0.8\textwidth]{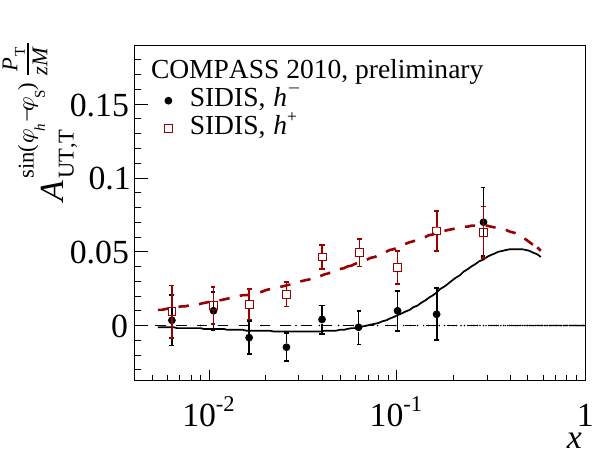}
\caption{\label{fig:AwSIDIS}
	The weighted Sivers asymmetry in SIDIS\,\cite{compass:2018weighted}, fitted. 
	Statistical errors only.}
\end{minipage}
\hspace{0.01\textwidth}
\begin{minipage}{0.48\textwidth}
    \includegraphics[width=0.8\textwidth]{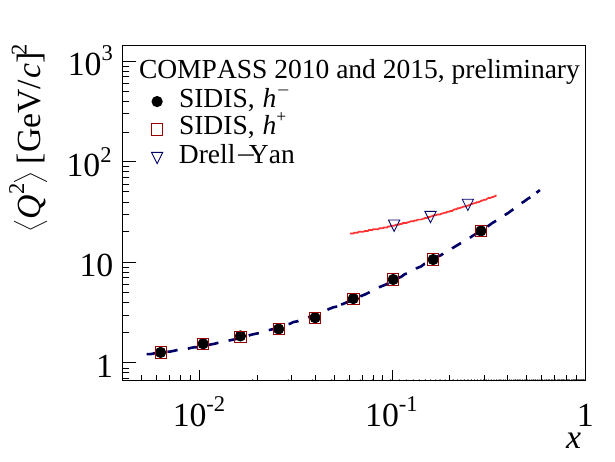}
    \caption{\label{fig:Qsq}
    	Mean $Q^2$ of the SIDIS and DY event samples, interpolated by polynomials.}
\end{minipage}
\end{figure}

\begin{figure}[t]
\begin{minipage}{0.32\textwidth}
    \includegraphics[width=\textwidth]{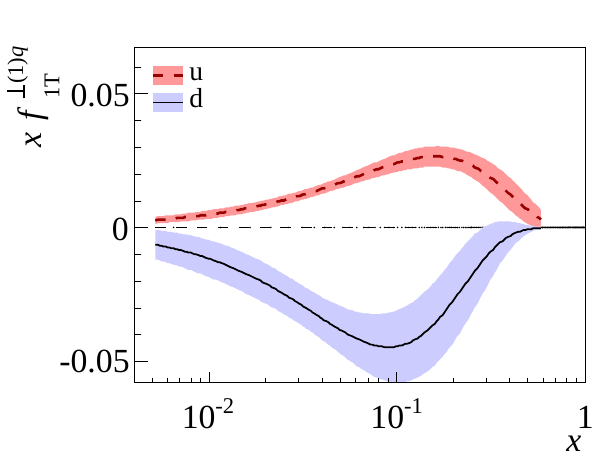}
    \caption{\label{fig:Sivers}
    	The first transverse moment of the Sivers function as a function of $x$ and 
    	$Q^2(x)$.}
\end{minipage}%
\hspace{0.01\textwidth}
\begin{minipage}{0.32\textwidth}
    \includegraphics[width=\textwidth]{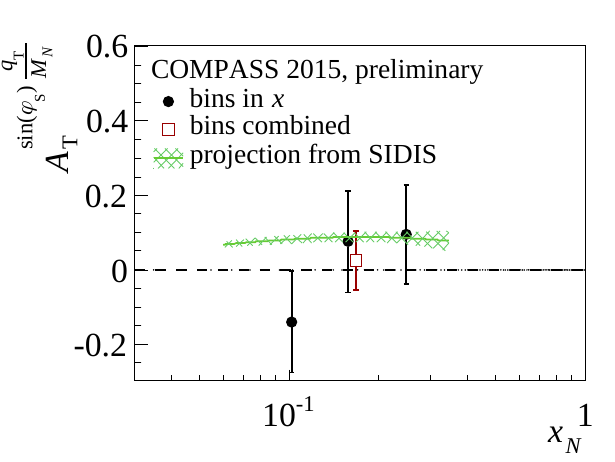}
    \caption{\label{fig:AwDY}
    	The weighted Sivers asymmetry in DY and the 
    	projection from SIDIS.}
\end{minipage}%
\hspace{0.01\textwidth}
\begin{minipage}{0.32\textwidth}
    \includegraphics[width=\textwidth]{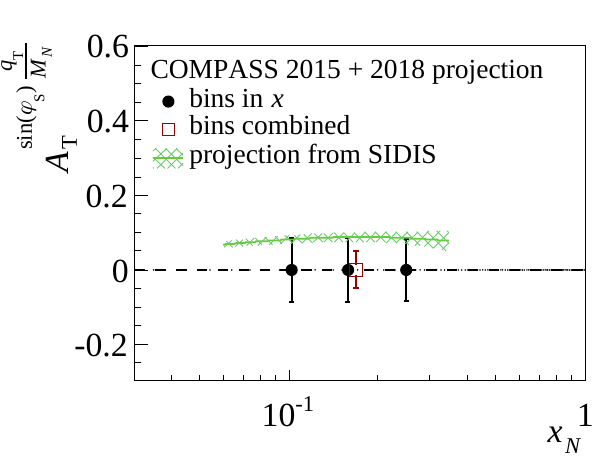}
    \caption{\label{fig:AwDYproj}
    	A projection for a combined analysis of 2015 and 2018 data.}
\end{minipage}%
\end{figure}

Taking the unpolarised PDFs from CTEQ~5D global fit\,\cite{cteq5,lhapdf6} and the FFs from DSS~07 global fit\,\cite{dss:2007}, the first transverse moments of u and d Sivers functions $\fSiv{(1)\mr{u/d}}(x)$ are the only unknowns. Unlike in Ref.\,\cite{compass:2018weighted}, here we utilized a parametrisation
$x \fSiv{(1)q}(x) = a_q \, x^{b_q} \, (1-x)^{c_q}$ and we fitted the weighted asymmetries for $h^+$ and $h^-$ (Fig.~\ref{fig:AwSIDIS}). 
In our kinematics, $x$ and $Q^2$ are correlated. Therefore we took the PDFs and FFs at the mean $Q^2$ corresponding to each $x$ bin (Fig.~\ref{fig:Qsq}). The result is shown in Fig.~\ref{fig:Sivers}. The $1\sigma$ error-bands in the figure account only for the uncertainty of the fit and for the statistical errors of the data. The statistical uncertainties of the PDFs and FFs have been neglected; a variation of their choice was observed to cause differences up to $2\sigma$.

The projection for DY can be obtained using \refeq{eq:siv}. Neglecting sea quarks in the flavour sums and assuming the change-of-sign prediction~\cite{collins:2002} one obtains
\beq
	\label{eq:sivapp}
	\A{T}{\sin\phiS\frac{\qTscal}{M_\mr{p}}}(x_\pi,x_N) 
		= -2 \frac{\fSivb{(1)\mr{u}}(x_N)}{f_{1,\mr{p}}^\mr{u}(x_N)}
		= 2 \frac{\fSivb{(1)\mr{u}}(x = x_N)|_\mr{SIDIS}}{f_{1,\mr{p}}^\mr{u}(x_N)}.
\eeq
The valence approximation is justified, as our experiment covers the valence region of both p and $\pi^-$; an additional suppression comes from the quark charge. The advantage is the cancellation of the pion PDF, which is not known very well. We used the same $f_{1,\mr{p}}^\mr{u}$ as in SIDIS, evaluated at the mean $Q^2$ of the DY sample (Fig.~\ref{fig:Qsq}). We neglected the evolution of the Sivers function between the SIDIS and DY $Q^2$. The result is compared to the measured asymmetries in Fig~\ref{fig:AwDY}. Again, only statistical errors are considered; the systematic uncertainty from the valence approximation was estimated to be smaller than $1\sigma$ using the GRV-PI PDF set\,\cite{grv:1992,lhapdf6}. The variation of FF set in the SIDIS fit has an impact up to $2\sigma$ in the lower $x_N$ region. As stated in Ref.\,\cite{compass:2017dy}, the significance of the result does not yet allow us to claim the validity of the `change-of-sign' rule. A second Drell--Yan run has been successfully finished in 2018. A projection for combined analysis of the two runs is shown in Fig.~\ref{fig:AwDYproj} assuming the statistics in 2018 to be 1.5 times larger than in 2015.

\section{Boer--Mulders function in weighted asymmetries in SIDIS and Drell--Yan}
\label{sec:boermulders}
The Boer--Mulders (BM) function is expected to change sign between SIDIS and DY as well. However, it will be more difficult to test it, as more TMDs are involved. The BM function of the proton can be accessed in our experiment via the spin-independent $\cos2\phi$ asymmetry or its weighted version, as proposed in Ref.~\cite{sissakian:2005b}. At leading twist, the weighted asymmetry is
\beq
    \A{U}{\cos2\phi\,\frac{\qTscal^2}{4 M_\pi M_\mr{p}}} 
	= 2 \frac { \sum_q e_q^2 \bigl[ h_{1,\pi^-}^{\perp (1) \bar{q}} (x_\pi)\,
					                h_{1,\mr{p}}^{\perp (1) q} (x_\mr{p})
					               + \qqswap \bigr] }
			  {\sum_q e_q^2  \bigl[ f_{1,\pi^-}^{\bar{q}}(x_\pi)\, 
						            f_{1,\mr{p}}^q(x_N) + \qqswap \bigr]}.
\eeq
Assuming the BM function of the sea to be zero, the numerator simplifies to the product of the first transverse moments of pion and proton BM functions,
$\frac{4}{9} \, h_{1,\pi^-}^{\perp (1) \bar{\mr{u}}} (x_\pi) \,
                h_{1,\mr{p}}^{\perp (1) \mr{u}} (x_\mr{p})$.

The pion BM function, which is very interesting by itself as the only nontrivial TMD PDF in pion apart from $f_{1\pi}$, can be independently obtained for the first time from the $\A{T}{\sin(2\phi-\phiS) \frac{\qTscal}{M_\pi}}$ asymmetry shown in Fig.~\ref{fig:wasym} using \refeq{eq:transv}. The u-quark transversity $h_{1,\mr{p}}^\mr{u}$, which is needed for this, is available from various extractions (for a comparison see e.g.\,\cite{radici:2015}).

The measurement of BM function in SIDIS, needed for the sign-change test, is difficult due of the mixing with the Cahn effect\,\cite{cahn:1978}. The extractions so far relied on strong assumptions\,\cite{barone:2009}. The new SIDIS data collected by COMPASS in 2016 and 2017\,\cite{moretti:2018} may help to improve the situation.
\section{Conclusions}
The transverse-momentum-weighted asymmetries offer an interesting alternative to the standard ones, since they rely on slightly different assumptions. They have a straightforward interpretation without the need of a model of the $\kT$-dependence of the TMD PDFs. We measured all three leading twist \qTscalt-weighted TSAs present in $\pi\mr{p}^\uparrow$ DY. Unlike the standard Sivers asymmetry\,\cite{compass:2017dy}, the weighted asymmetry is compatible, within $1\sigma$, with both sign-change and no-sign-change hypotheses for the Sivers function. The new DY data collected in 2018 will provide a better comparison.

%% file: Matousek_SPIN18.bbl
\providecommand{\href}[2]{#2}\begingroup\raggedright\endgroup